\documentstyle[12pt]{article}
\begin{document}
\setlength{\baselineskip}{0.8 cm}

\hfill December, 1996

\vskip 0.5in
\centerline{\Large Resonant Light Scattering to Measure BEC-Pairing}
\vskip 0.25in
\centerline{ Eddy Timmermans and Paolo Tommasini}
\vskip 0.15in
\centerline{Institute for Theoretical Atomic and Molecular Physics}
\centerline{Harvard-Smithsonian Center for Astrophysics}
\centerline{Cambridge, MA 02138}
\vskip 0.5in

\centerline{ABSTRACT}
\vskip 0.20in

We present a single-scattering formalism for incoherent 
resonant light scattering
by dilute quantum gas systems such as the atomic-trap Bose-Einstein
condensates.  We show that resonant scattering gives access to more
information than the dynamical structure factor, familiar from non-resonant
scattering.  In particular,
we show that the detuning dependence of the incoherent scattering cross-section
allows the direct determination of the BEC pairing density $\langle \psi \psi
\rangle$, which is a broken symmetry and provides evidence that the condensate
is not in a good number state.

\newpage

	The unusual properties of the atomic-trap Bose-Einstein condensates
\cite{Ket}--\cite{Hul}
make them prime examples of dilute many-body systems 
with highly interesting
microscopic structures.  
Experimentally, the most convenient probe
is resonant optical scattering.  
From the theoretical perspective, 
this poses an interesting problem -- resonant scattering
is a second-order process and cannot be described by means of the usual
Van Hove theory \cite{VH}.
Nevertheless, as Javaneinen pointed out
\cite{Jav1} \cite{Jav2}, a Van-Hove-like expression is recovered in 
the off-resonant limit.  In this paper, we discuss a single scattering
formalism that is valid for arbitrary values of the
detuning.

Typically, the width $\gamma$ of the resonant transition is much larger
than the trap frequency $\omega_T$ (twice the ground state energy of the
trapping potential, $\hbar = 1$ in our units).  Nevertheless, 
for some long-lived states, $\gamma$ can be comparable to the relevant 
excitation energies of the many-atom system, so that the scattering is fast
on the scale of the excited atom motion, but not necessarily on the scale
of the many-body dynamics.  In that case, we find that
resonant scattering gives 
access to 
more information about the many-body structure than the dynamical structure
factor.  In particular, for a dilute Bose condensate we are lead to the
remarkable conclusion that resonant scattering allows a direct determination
of the pairing density  $\langle \psi \psi \rangle$, related to the 
Bose symmetry breaking.

	It is instructive to recall the non-resonant scattering result.
We adopt the convention that for an incident particle of momentum 
${\bf k}_{\rm{in}}$ and frequency 
$\omega_{\rm{in}}$, scattered into a state of momentum ${\bf k}_{\rm{out}}$ and
frequency $\omega_{\rm{out}}$, ${\bf q}$ is
the momentum transferred 
to the target system, ${\bf q} = 
{\bf k}_{\rm{in}} - {\bf k}_{\rm{out}}$, and 
$\omega = \omega_{\rm{in}} - \omega_{\rm{out}}$.
Van Hove showed that the differential cross section $d^{2} \sigma /
d \Omega \; d\omega$, where $d\Omega$ is an infinitesimal solid angle,
only depends on ${\bf q}$ and $\omega$, and is equal to
\begin{equation}
\frac{d^{2} \sigma}{d \Omega\; d\omega}
= |f({\bf q})|^{2}  S({\bf q},\omega)\; ,
\label{e:lvh}
\end{equation}
\noindent
where $f({\bf q})$ is the scattering 
length describing the scattering
of an incident particle by an individual target particle, 
and $S({\bf q},\omega)$ is the
dynamical structure factor of the many-body system.  
The structure factor
is the Fourier-transform of the density-density correlation function,
\begin{equation}
S(q) = (2 \pi)^{-1} \int d^{3} x \; d^{4} x' \exp[i q \cdot 
(x-x') ] \; \langle \hat{\rho} (x') \hat{\rho} (x) \rangle \; \; ,
\label{e:sf} 
\end{equation}
\noindent
where $\langle \; \rangle$ denotes the thermally
averaged expectation value over the initial states of the 
many-body target system, 
and where we introduced the four-vector notation,
$q \equiv ({\bf q},\omega)$, $x \equiv ({\bf x},t)$, and $q \cdot x =
{\bf q} \cdot {\bf x} - \omega t$.  In second quantization, the density 
operator is equal to : 
\begin{equation}
\hat{\rho} (x) = \hat{\psi}^{\dagger} (x) \hat{\psi} (x) \; ,
\label{e:denso}
\end{equation}
where $\hat{\psi}$ and $\hat{\psi}^{\dagger}$ are the annihilation and creation
fields in the Heisenberg picture. 
The integration $\int d^{3} x$ in (\ref{e:sf})
is over the three spatial components, the time component being
fixed. 

	Unlike non-resonant scattering, 	
resonant photon scattering involves two interactions -- 
the atomic excitation
while absorbing the incident photon, and the
de-excitation accompanied by photon emission.
In summing over scattering histories, we integrate over 
the photon absorption amplitude at $x_{1}$, the photon emission amplitude
at $x_{2}$, and the amplitude for the excited atom to 
move from $x_{1}$ to $x_{2}$, which is the excited atom propagator 
$G(x_{2},x_{1})$, where $x_{1} \neq x_{2}$ \cite{rwa}. 
To deal with this type of non-locality,
it is useful to work in the Wigner-representation, 
where a two-point function, $f(x_{1},x_{2})$ in coordinate representation,
is represented by
\begin{equation}
f_{\rm{W}} (x;p) 
= \int d^{4} r \; f(x+r/2,x-r/2) \exp[ip \cdot r] 
\;. 
\label{e:wr}
\end{equation}
\noindent
Similarly, we introduce the Wigner distribution operator:
\begin{equation}
\hat{\rho}_{W} (x;p) = (2 \pi)^{-4} \int d^{4} r \; \hat{\psi}^{\dagger} (x+r/2)
\hat{\psi} (x-r/2) \exp[ip \cdot r] \;, 
\label{e:wdo}
\end{equation}
\noindent
the expectation value of which gives the Wigner
distribution \cite{Wig}, \cite{Lang},
the quantum analogue of the classical phase-space
distribution function.
A straightforward but somewhat lengthy derivation yields the following
cross-section for 
light scattering involving an atomic transition of natural frequency 
$\omega_{0}$ and dipole moment ${\bf d}$ :
\begin{eqnarray}
 &&\frac{d^{2} \sigma \left[ \hat{\epsilon}_{\rm{in}},\hat{\epsilon}_{\rm{out}};
\Delta \right]}{d \Omega \; d(\hbar \omega)}
= \left( \frac{ 3 \gamma}{4 k}\right)^{2} \left|
(\hat{\epsilon}_{\rm{in}} \cdot \hat{\bf d} ) \;
(\hat{\epsilon}_{\rm{out}} \cdot \hat{\bf d} ) \right| ^{2} 
\frac{1}{2 \pi} \times 
\int d^{3} x \; d^{4} x' \; d^{4} p \; d^{4} p' \;
\nonumber\\
&&\exp \left[ i q \cdot (x-x') \right]
G_{W}(x;p+k_{+}) \; G_{W}^{\ast}(x';p'+k_{+}) \;
\langle \hat{\rho}_{W} (x',p') \hat{\rho}_{W}(x,p) \rangle  \; ,
\label{e:crg}
\end{eqnarray}
\noindent
where $\Delta$ is the detuning of the incident photons,
$\Delta = \omega_{\rm{in}}-\omega_{0}$,
$k_{+}$ is the average of the incident and outgoing photon momentum
four-vector, $k_{+} = ({\bf k}_{+}, \omega_{+})$ $\equiv 
( \left[ {\bf k}_{\rm{in}} + {\bf k}_{\rm{out}} \right] /2, 
\left[ \omega_{\rm{in}} + \omega_{\rm{out}} \right] /2 )$, 
$\hat{\bf d} = {\bf d} / |{\bf d}|$ 
and 
$\hat{\epsilon}_{\rm{in}}$ and $\hat{\epsilon}_{\rm{out}}$ 
are the polarization directions
of the incident and detected photons.

	The propagator G can be computed from the eigenstates
of the effective potential $\tilde{V}({\bf x})$
experienced by the excited atoms.  Nevertheless, 
if the excited state lifetime is so short that the excited atom
experiences a change in $\tilde{V}$ small compared to the width $\gamma$,
(i.e. $| \; \left[ {\bf F} \cdot {\bf v} \right]  / \gamma^{2} \; | \ll 1$, 
where ${\bf F}$
is the force and ${\bf v}$ the velocity of the excited atom), 
then $G_{W}$ is approximated accurately by the homogeneous propagator with
$\tilde{V}({\bf x})$ as position dependent shift: 
\begin{eqnarray}
G_{W}({\bf x}; p+k_{+}) &=& \frac{1}
{p_{0} + \omega_{+}
- \left[ H_{\rm{e}} ({\bf p}+{\bf k}_{+},{\bf x}) + \omega_{0} \right]
} \nonumber\\
&=& \frac{1}{\Delta -
\left[ H_{\rm{e}} ({\bf p}+{\bf k}_{+},{\bf x})
- (p_{0}-\omega/2) \right]} \;, 
\label{e:prop}
\end{eqnarray}
where $p_{0}$ is the frequency-component of the momentum four-vector p, 
and $H_{e}$ is the self-energy of the excited atom, including 
the potential $\tilde{V}$ and the width $-i \gamma/2$.
Expanding the propagator $G_{W}$ in powers of the inverse detuning,
gives a $\Delta^{-1}$--expansion of the cross-section,
with coefficients that
are n-th order moments of the $\left[ H_{\rm{e}}({\bf p}+{\bf k}_{+};
{\bf x}) - (p_{0}-\omega/2) \right]$--functions of 
the propagators (\ref{e:prop}) with
respect to the $\langle \hat{\rho}(x';p') 
\hat{\rho}(x;p) \rangle$--correlation function. 
At first sight, the appearance of ${\bf k}_{+}$, instead of the momentum
of the excited atom, might seem puzzling.  The problem is resolved when 
we return to the ordinary coordinate
representation using
\begin{eqnarray}
\int d^{4} x \; d^{4} p \; \exp[i q \cdot x] \; F(p)\; \hat{\rho}_{W} (x;p)
&=& \int d^{4} x \; \exp[i q \cdot x] \; \hat{\psi}^{\dagger} (x)
\; \stackrel{\rightharpoonup}{\cal{F}} (\hat{p} + q/2) \; \hat{\psi} (x) 
\nonumber\\
&=&  \int d^{4} x \; \exp[i q \cdot x] \; \hat{\psi}^{\dagger} (x)
\; \stackrel{\leftharpoonup}{\cal{F}} (\hat{p} - q/2) \; \hat{\psi} (x) \; ,
\label{e:ret}
\end{eqnarray}
\noindent
where F(p) is an arbitrary function, and 
$\cal{F}$($\hat{p}$) the operator obtained by replacing the p-vector
in the expression for F, by the p-operator, $\hat{p}_{j} = \frac{1}{i} 
\partial / \partial x_{j}$, if j=1,2 or 3, and $\hat{p}_{0} = i 
\partial / \partial t$.
The right $\rightharpoonup$ and left $\leftharpoonup$ arrows indicate that
the p-operator only acts upon the field operator immediately
to the right or left.
Since  ${\bf q}/2 + {\bf k}_{+} = {\bf k}_{\rm{in}}$, the
$\left[ H_{\rm{e}}({\bf p}+{\bf k}_{+};
{\bf x}) - (p_{0}-\omega/2) \right]$-functions give
an operator ${\cal{H}}_{\rm{e}}({\bf k}_{\rm{in}}+\hat{\bf p} ; {\bf x})
-i \partial /\partial t$. 
The time derivative $i \partial / \partial t$, 
acts on a `hole' state.
In this sense, ${\cal{H}}_{\rm{e}}
- i \partial / \partial t$ 
describes the evolution of the excited atom-hole
pair and we denote the operator ${\cal{H}}_{\rm{e}}
- i \partial / \partial t$ by ${\cal{H}}_{\rm{e-g}}$.
The cross-section is then equal to
\begin{eqnarray}
\frac{d^{2} \sigma \left[ \hat{\epsilon}_{\rm{in}},\hat{\epsilon}_{\rm{out}};
\Delta \right]}{d \Omega \; d(\hbar \omega)}
= \left|  (3 \gamma /4k) (\hat{\epsilon}_{\rm{in}} \cdot \hat{\bf d})
(\hat{\epsilon}_{\rm{out}} \cdot \hat{\bf d})^{\ast}\right|^{2}
\frac{1}{2\pi}\; \times \; \int d^{3} x \; d^{4} x' \exp \left[ i q \cdot
(x-x') \right] 
\nonumber\\
\langle \hat{\psi}^{\dagger} (x') \left[
\frac{1}{\Delta-\stackrel{\leftharpoonup}
{\cal{H}}_{\rm{e-g}}(x')} \right]^{\dagger} \hat{\psi} (x')
\hat{\psi}^{\dagger}(x)  \left[
\frac{1}{\Delta-\stackrel{\rightharpoonup}
{\cal{H}}_{\rm{e-g}}(x')} \right] \hat{\psi} (x) \rangle \; .
\label{e:comp}
\end{eqnarray}
\noindent
For large detunings, 
we approximate the denominators
in (\ref{e:comp}) by $\Delta$, giving the off-resonant limit 
reported by Javaneinen \cite{Jav1} : 
\begin{equation}
\lim_{\Delta \rightarrow \infty}
\frac{d^{2} \sigma \left[ \hat{\epsilon}_{\rm{in}},\hat{\epsilon}_{\rm{out}};
\Delta \right]}{d \Omega \; d(\hbar \omega)}
= \left| \frac{ (3 \gamma /4k) (\hat{\epsilon}_{\rm{in}} \cdot \hat{\bf d})
(\hat{\epsilon}_{\rm{out}} \cdot \hat{\bf d})^{\ast}}{\Delta} \right|^{2}
S({\bf q},\omega) \; , 
\label{e:jav}
\end{equation}
\noindent
where we recognize 
the off-resonant limit of the scattering length for optical 
resonant scattering from atoms, showing the similarity of (\ref{e:jav}) 
to the Van-Hove expression (\ref{e:lvh}).

The general result (\ref{e:comp}) 
indicates how and when the off-resonant
limit (\ref{e:jav}) breaks down.  
Especially for cold fermionic atom systems, it can be important to
correctly include the effects of the Doppler shifts, 
contained in (\ref{e:comp}). For the atomic Bose condensates, the recoil 
energies can be of importance, as we show below.

	A straightforward 
generalization of the customary detailed balance argument \cite{Pines} for
the structure factor shows that changing the sign of the energy transfer
$\omega \rightarrow -\omega$ (while keeping $\Delta$ constant) similarly
gives :
\begin{equation}
\frac{d^{2} \sigma}{d\Omega \; d \omega} (\Delta ;\omega)
= \; \exp( \beta \omega ) \; 
\frac{d^{2} \sigma}{d\Omega \; d \omega} (\Delta - \omega ; -\omega) \; , 
\label{e:db}
\end{equation}
\noindent
where $\beta$ is the inverse temperature.
Verifying (\ref{e:db}) experimentally (which does note require angular
resolution) for the atomic-trap BEC systems can test the hypothesis
that the observed condensate is in thermal equilibrium and
provides a direct measurement of the temperature.

	We now discuss scattering from a dilute homogeneous BEC at zero 
temperature.
Since our concern is not with coherent scattering (observed at 
$\omega$ = 0 , ${\bf q}$ = 0), we calculate the incoherent
part of the cross-section, $d^{2} \sigma_{\rm{nc}} / d\Omega d\omega$,
subtracting $\langle \hat{\psi}^{\dagger} \hat{\psi} \rangle
\langle \hat{\psi}^{\dagger} \hat{\psi} \rangle$ from
$\langle \hat{\psi}^{\dagger} \hat{\psi} 
\hat{\psi}^{\dagger} \hat{\psi} \rangle$ in (\ref{e:crg}).  We expand
the $\hat{\psi}$-fields in plane wave states, $\hat{\psi}(x)
= V^{-1/2} \sum_{\bf k} a_{\bf k} (t) \exp\left[ i{\bf k} \cdot {\bf x} \right] 
$, $\hat{\psi}^{\dagger}(x)
= V^{-1/2} \sum_{\bf k} a^{\dagger}_{\bf k} (t)
\exp\left[ -i{\bf k} \cdot {\bf x} \right]$,
where $V$ is the volume ($V \rightarrow \infty$
in the end),
and make the mean-field Bogoliubov approximation, treating $a_{{\bf k}=0},
a^{\dagger}_{{\bf k}=0}$ as c-numbers, $a_{{\bf k}=0},
a^{\dagger}_{{\bf k}=0} \rightarrow \sqrt{N_{0}}$, keeping terms up
to order $N_{0}$.  The result is 
\begin{eqnarray}
\frac{d^{2} \sigma_{\rm{nc}}}{d\Omega \; d\omega}
\left[ \hat{\epsilon}_{\rm{in}},\hat{\epsilon}_{\rm{out}};
\Delta \right]
&\approx&
N_{0} \; 
\left|(3 \gamma/4 k)
(\hat{\epsilon}_{\rm{in}} \cdot \hat{\bf d} ) \;
(\hat{\epsilon}_{\rm{out}} \cdot \hat{\bf d} ) \right| ^{2}
\; \; \delta(\omega - \omega_{\bf q}) \times 
\nonumber\\
&&\left( \left| \frac{1}{\stackrel{-}{\Delta} + i \gamma/2 - \omega_{\bf q} }
\right| ^{2} \langle a^{\dagger}_{-{\bf q}} a_{-{\bf q}} \rangle \; +
\left| \frac{1}{\stackrel{-}{\Delta} + i \gamma/2} \right| ^{2}
\langle a_{\bf q} a^{\dagger}_{\bf q} \rangle \right.
\nonumber\\
&&
+ \left[ \frac{1}{\stackrel{-}{\Delta} - i \gamma/2 - \omega_{\bf q} } \right]
\left[ \frac{1}{\stackrel{-}{\Delta} + i \gamma/2} \right]
\langle a^{\dagger}_{-{\bf q}} a^{\dagger}_{\bf q} \rangle
\nonumber\\
&&+ \left. 
\left[ \frac{1}{\stackrel{-}{\Delta} + i \gamma/2 - \omega_{\bf q}} \right]
\left[ \frac{1}{\stackrel{-}{\Delta} - i \gamma/2} \right]  
\langle a_{\bf q} a_{-{\bf q}} \rangle \right) ,
\label{e:snc}
\end{eqnarray}
\noindent
where $\stackrel{-}{\Delta}$ is the effective detuning, $\stackrel{-}
{\Delta} = \Delta - \tilde{\omega}(|{\bf k}|) - \tilde{V}$, 
and $\tilde{\omega}(|{\bf k}|)$ represents the kinetic energy of an excited 
atom with momentum equal to the resonant wave number.
In addition to the peak at $\omega = \omega_{\bf q}$ represented by 
(\ref{e:snc}), the incoherent scattering spectrum at fixed scattering angle
and detuning, has a continuous background, caused by  
${\bf k}$-modes (${\bf k} \neq 0$) excited 
to ${\bf k} + {\bf q}$ states.  In most cases of interest, the ratio 
of the integrated background
intensity to the peak intensity is of the order
of the depletion $\left( N - N_{0} \right) / N_{0}$, where N is the total 
number of atoms.

The Bogoliubov approximation implies that in each process a particle is
taken out of, or put into the condensate.  
This leaves only two possibilities if ${\bf q}$ is the momentum transfer: 
a particle
is excited from the condensate and ends up into a final state of momentum
$+ {\bf q}$ (corresponding to the subscript ${\bf q}$
in (\ref{e:snc})), or a particle is taken from the initial state of
momentum $-{\bf q}$ 
and ends up in the condensate (corresponding to the 
$-{\bf q}$-subscripts).
In the $+{\bf q}$-process,
the excited atom has momentum ${\bf k}_{\rm{in}}$,
in the $-{\bf q}$ process, the excited atom has momentum $- {\bf q} +
{\bf k}_{\rm{in}} = {\bf k}_{\rm{out}}$.  In either case, the excited atom
has a center-of-mass energy that is $\tilde{\omega}(|{\bf k}|)+\tilde{V}$.
On the other hand, the initial-state energy for the ${\bf q}$ and $-{\bf q}$ 
processes are different, so that the 
$\cal{H}_{\rm{e-g}}$-operator gives 
$\tilde{\omega}(|{\bf k}|) + \tilde{V} - i \gamma/2$ for
the $+ {\bf q}$-process and $\tilde{\omega}(|{\bf k}|) +
\tilde{V} - \omega_{\bf q} - i\gamma/2$ for
the $- {\bf q}$ scattering.
The $-{\bf q}$ process leaves the condensate with an extra particle,
whereas in the $+{\bf q}$ process, a particle is removed from the
condensate.  In the usual single particle picture, one could expect these
processes to give orthogonal final states, thereby precluding any interference.
Nonetheless, the condensate is expected to be in a coherent state, rather
than a number state, and the $+{\bf q}$ and $-{\bf q}$ final states have
a finite overlap proportional to $\langle a_{\bf q} a_{-{\bf q}} \rangle$.
Conversely, detecting the interference of the $+{\bf q}$ and $-{\bf q}$
scattering events answers the question wether or not the condensate
is in a good number state. 
With regards to this isssue, 
which has received considerable attention in the recent
literature \cite{Javyoo}--\cite{wong2},
we note that the incoherent scattering scheme provides
an alternative to the interfering condensate experiments.

To experimentally detect the interference, we note that the difference
in recoil energy 
for the $+{\bf q}$ and $-{\bf q}$ scattering processes gives
a distinct detuning dependence to the interference contribution in the cross
section (\ref{e:snc}):
\begin{eqnarray}
&&
\frac{d^{2} \sigma_{\rm{nc}}}{d\Omega \; d\omega}
\left[ \hat{\epsilon}_{\rm{in}},\hat{\epsilon}_{\rm{out}};
\Delta \right]
\approx
N_{0} \;
\left|(3 \gamma/4 k)
(\hat{\epsilon}_{\rm{in}} \cdot \hat{\bf d} ) \;
(\hat{\epsilon}_{\rm{out}} \cdot \hat{\bf d} ) \right| ^{2}
\; \; \delta(\omega - \omega_{\bf q}) \times
\nonumber\\
&&
\left(
\frac{1}{\left(\stackrel{-}{\Delta} - \omega_{\bf q} \right) ^{2} + \left(
\gamma /2 \right) ^{2} }
\left[ \langle  a^{\dagger}_{-{\bf q}} a_{-{\bf q}} \rangle +
\langle a_{\bf q} a_{-{\bf q}} \rangle
\right] +
\frac{1}{\left(\stackrel{-}{\Delta} \right) ^{2} + \left(
\gamma /2 \right) ^{2} }
\left[ 1+ \langle  a^{\dagger}_{\bf q} a_{\bf q} \rangle +
\langle a_{\bf q} a_{-{\bf q}} \rangle
\right]
\right.
\nonumber\\
&&
\left.
- \frac{ \omega_{\bf q}^{2} }
{\left( \stackrel{-}{\Delta} - \omega_{\bf q} \right) ^{2} + \left(
\gamma /2 \right) ^{2} } \times
\frac{1}{\left(\stackrel{-}{\Delta} \right) ^{2} + \left(
\gamma /2 \right) ^{2} }
\langle a_{\bf q} a_{-{\bf q}} \rangle
\right) \; ,
\label{e:fin}
\end{eqnarray}
\noindent
where we used that
$\langle a^{\dagger}_{-{\bf q}} a_{-{\bf q}} \rangle = 
\langle a^{\dagger}_{\bf q} a_{\bf q} \rangle$,
$\langle a_{\bf q} a^{\dagger}_{\bf q} \rangle = 1 + 
\langle a^{\dagger}_{\bf q} a_{\bf q} \rangle$ and
$\langle a^{\dagger}_{-{\bf q}} a^{\dagger}_{\bf q} 
\rangle$ = $\langle a_{{\bf q}} a_{-{\bf q}} \rangle$.  
Thus, the detuning dependence of the peak intensity differs 
from the simple Lorentzian 
$\left[ \stackrel{-}{\Delta}^{2} + (\gamma/2)^{2} \right]^{-1}$.
The simplicity of the actual detuning dependence, which contains 
only two parameters, the occupation number $\langle a^{\dagger}_{\bf q}
a_{\bf q} \rangle$ and pairing matrix element 
$\langle a_{-{\bf q}} a_{\bf q} \rangle$, suggests a simple fitting of
the experimental curves to determine their values.
In Fig.(1) we show the detuning dependence of the intensity in the 
peak of the incoherent scattering spectrum, for the special case that
the scattering angle corresponds to a momentum transfer for which the
excitation energy $\omega_{\bf q}$ is equal to the chemical potential
$\mu$ and $\mu$ is equal to the width $\gamma$. The full curve
shows the actual intensity, whereas the dotted lines indicate the contributions
proportional to the occupation number and the pairing matrix element.  The
occupation number and pairing matrix elements where calculated in the 
Bogoliubov approximation at T=0 (the finite temperature generalization
is straightforward, see for example \cite{Walls}).
Requiring the non-Lorentzian contribution 
to be measurable leads to the condition that $\omega_{\bf q}$ is 
of the order of 
$\gamma$.  The Bogoliubov theory at T=0 gives $\langle
a_{\bf q} a_{-{\bf q}} \rangle = -\mu / 2 \omega_{\bf q}$, yielding a magnitude
of the non-Lorentzian term relative to the other contributions that is of the
order of $\sim (\frac{\omega_{\bf q}}{\gamma})  ( \frac{\mu}{\gamma/2})$.
Therefore, the best signal is obtained for backscattering, 
$ q = 2k$, $\omega_{\bf q}
\sim 4 \omega_{r}$, where $\omega_{r}$ is the recoil energy, with a 
relative magnitude of $\sim 2 \;  (\frac{\omega_{r}}{\gamma/2} )
\; (\frac{\mu}{\gamma/2})$.

Of course the atomic-trap system is not homogeneous. 
In fact, we cannot require the condensate to be too close to homogeneity
because if 
the system is optically thick, the single scattering approximation
breaks down.  Nevertheless, it is possible to have an optically thin system
for which 
$\mu \gg \omega_{T}$, in which case the Thomas-Fermi description
is valid and the condensate \cite{Leg}, \cite{Kag}, as well as the fluctuations 
\cite{us}, \cite{Grif} behave locally in the same manner as the homogeneous
system. In a
Thomas-Fermi description of the incoherent scattering, the single peak is
`broadened', giving a feature in the frequency interval from 
$\sqrt{(q^{2}/2m+\mu)^{2} - \mu^{2}}$ to $q^{2}/2m$.  
Within this feature, the intensity of a frequency interval $\omega'$ to
$\omega' + d \omega$ has information about the spatial region in which
the local excitation with momentum ${\bf q}$ has an excitation energy 
$\omega_{\bf q}({\bf r})$ in the $d \omega$ - interval, $\omega' < 
\omega_{\bf q}({\bf r}) < \omega'+d \omega$.  If the density is nearly
constant in this spatial region, 
the detuning dependence of the intensity in the $d\omega$
- interval is described by the above theory.

	In conclusion, we have shown that resonant light scattering can 
measure the pairing matrix elements of Bose-Einstein condensates.
The finite value of the `pairing' density, $\langle
\psi \psi \rangle$, is an example of a higher-order broken symmetry and
is of fundamental interest.  Measuring its value would constitute a 
detailed test of mean-field theories and finding a non-zero result would
prove experimentally that the condensate is not in a number state. 

	 P.T. was supported by Conselho Nacional de
Desenvolvimento Cientifico e Tecnologico (CNPq), Brazil.
The work of E.T. is supported by the NSF through a grant for the
Institute for Atomic and Molecular Physics at Harvard University
and Smithsonian Astrophysical Observatory.

\newpage

\underline{\Large Figure Caption}
\\
\\
\\
Figure 1: Plot of the detuning dependence of the peak intensity in the
incoherent scattering spectrum.  The calculation was performed in
the Bogoliubov approximation at zero temperature for the special case that
the momentum transfer corresponds to an excitation energy 
$\omega_{\bf q}$ equal to the chemical potential
$\mu$, and $\mu$ is equal to the excited state width $\gamma$. The full curve
shows the actual intensity, the dotted line with negative values
shows the contribution proportional to the pairing number 
$\langle a_{{\bf q}} a_{-{\bf q}} \rangle$ (which is
negative).  The dotted line with the positive values represents the
contribution
proportional to the occupation number $\langle a_{{\bf q}}^{\dagger}
a_{\bf q} \rangle$.


\begin{thebibliography}{99}
\bibitem{rwa} In this formalism, we assume that the photon emission occurs
after the photon absorption, so that we work in the rotating wave approximation.
\end{thebibliography}
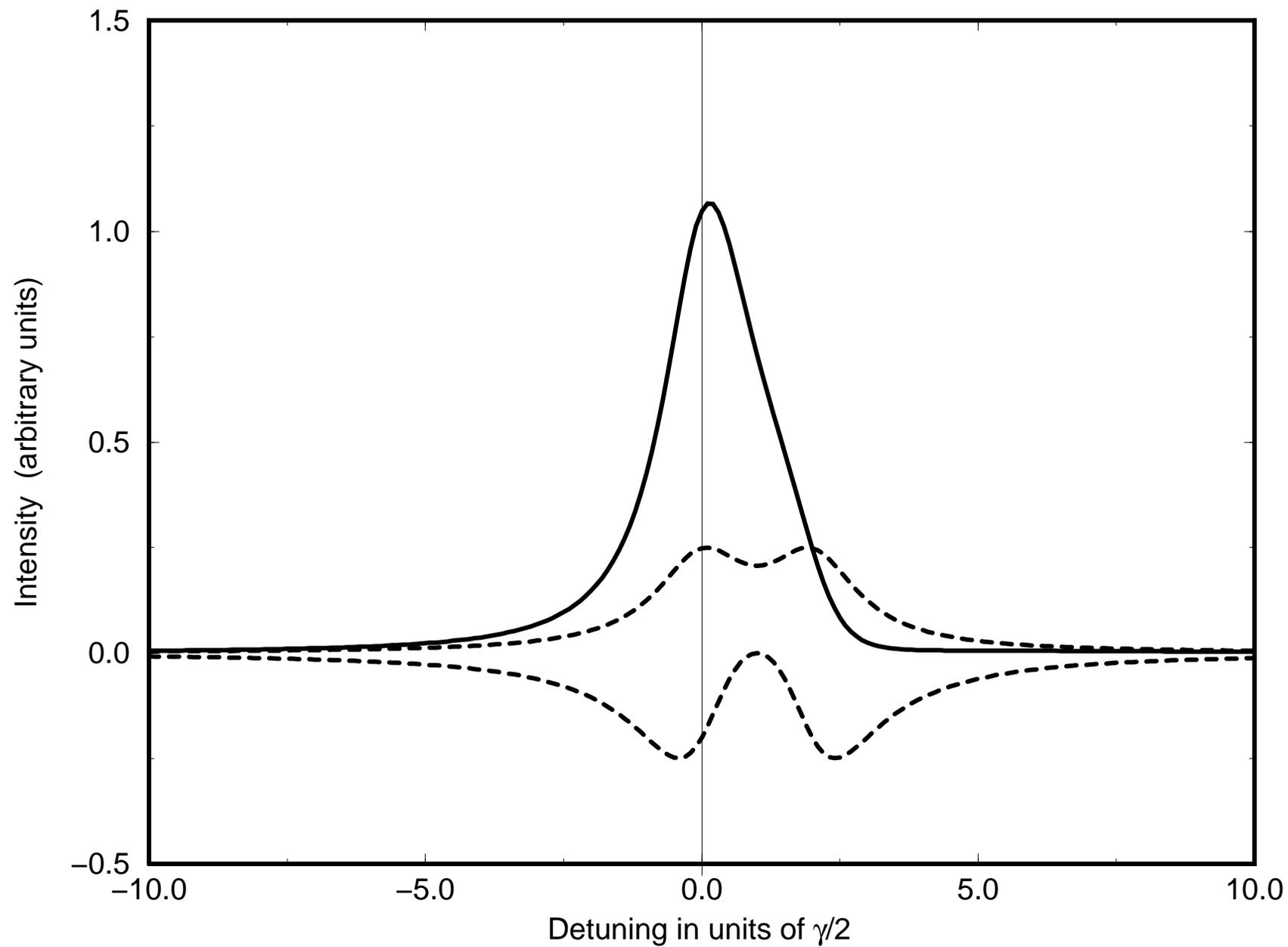
\end{document}